\documentclass[reprint,
 amsmath,amssymb,
 aps,
pra,
]{revtex4-2}

\usepackage{graphicx}
\usepackage{dcolumn}
\usepackage{bm}

\begin{document}

\preprint{APS/123-QED}

\title{Conformational and static properties of tagged chains in solvents: effect of chain connectivity in solvent molecules}

\author{Hong-Yao Li}
 \affiliation{School of Physical Science and Technology, Southwest University, Chongqing 400715, China}
\affiliation{Chongqing Key Laboratory of Micro-Nano Structure Optoelectronics, Chongqing 400715, China}
\author{Bokai Zhang}
\email{zbk329@swu.edu.cn}
 \affiliation{School of Physical Science and Technology, Southwest University, Chongqing 400715, China}
\affiliation{Chongqing Key Laboratory of Micro-Nano Structure Optoelectronics, Chongqing 400715, China}
\author{Zhi-Yong Wang}
\email{zywang@swu.edu.cn}
 \affiliation{School of Physical Science and Technology, Southwest University, Chongqing 400715, China}
\affiliation{Chongqing Key Laboratory of Micro-Nano Structure Optoelectronics, Chongqing 400715, China}
\date{\today}

\begin{abstract}
Polymer chains immersed in different solvent molecules exhibit diverse properties due to multiple spatiotemporal scales and complex interactions.  { Using molecular dynamics simulations, we study the conformational and static properties of tagged chains in different solvent molecules.  Two types of solvent molecules were examined: one type consisted of chain molecules connected by bonds, while the other type consisted of individual bead molecules without any bonds.  The only difference between the two solvent molecules lay in the chain connectivity.}  Our results show a compression of the tagged chains with the addition of bead or chain molecules.  Chain molecule confinement induces a stronger compression compared to bead molecule confinement.  In chain solvent molecules, the tagged chain's radius of gyration reached a minimum at a monomer volume fraction of $\sim0.3$.  Notably, the probability distributions of chain size remain unchanged at different solvent densities, irrespective of whether the solvent consists of beads or polymers.  Furthermore, as solvent density increases, a crossover from a unimodal to a bimodal distribution of bond angles is observed, indicating the presence of both compressed and expanded regions within the chain.  The effective monomer-solvent interaction is obtained by calculating the partial radial distribution function and the potential of the mean force.  In chain solvent, the correlation hole effect results in a reduced number of nearest neighbors around tagged monomers compared to bead solvents.  The calculation of pore size distribution reveals that the solvent nonhomogeneity induced by chain connectivity leads to a broader distribution of pore sizes and larger pore dimensions at low volume fractions. These findings provide a deeper understanding of the conformational behavior of polymer chains in different solvent environments.
\end{abstract}

\maketitle


\section{Introduction}
Understanding the conformational properties of polymers in complex fluids is of great importance in polymer physics \cite{Doi1988,McLeish2002}, and possesses extensive applications in cellular biology, and engineering materials.  In crowded cellular environments, the structure and dynamics of biopolymers have a profound impact on the essential cellular structures and functions, such as gene regulation, cell growth, and senescence. \cite{bio-Minton2020,bio-Janssen2023PRL,bio-Tabaka2013NAR,bio-Baptiste2022Nphys}  In engineering materials, doped nanoparticle (NP) fillers have been found to remarkably enhance various properties.  Unraveling the change of the conformation and motion of the polymer chains in nanocomposites is a crucial step toward understanding why doping strongly modifies material properties.\cite{Yan2013,Karatrantos2016,Kumar2017,Winey2015,Winey2020,Yang2022}

Small-angle neutron scattering (SANS) experiments revealed that the compression and expansion of chains depend on the ratio of the size of chains to that of NPs.  Tuteja et al. found that adding 10\% crosslinked polystyrene (PS) NP with a size ratio $R_g/R_n=1.6-5.7$ to an entangled linear PS melt led to approximately 20\% elongation of the PS chains.  \cite{Tuteja2008} Moreover, at higher adding NP volume fraction ($\phi=0.4$) and larger size ratios ($R_g/R_n=6-8$), experimental results discovered an impressive elongation of the chain dimension up to 60\%. \cite{Nakatani2001} There are also experimental observation that provide contradictory results.  For instance, in the PS/silica nanocomposites, the radius of gyration of chain was found to be unperturbed, and its distribution remained Gaussian statistics. \cite{Schweizer2007PRL} By combining SANS with small-angle X-ray scattering (SAXS) measurements and transmission electronic microscopy (TEM) images, Jouault et al. concluded that the chain conformation in PS/silica system was independent of the NP fillers. \cite{Boue2010Ma} Other independent SANS experiment uncovered that in a mixture of silica and poly(ethylene-propylene), the presence of silica NPs had no effect on the length of short chain matrix and resulted in a decrease of 12\% in the chain radius of gyration for long chain matrix
when the volume fraction of silica NPs reached 50\%. \cite{Nusser2010}

Theoretical and simulation methods are widely used to study the conformational changes in the chain surrounding NPs, as well as elucidate the role of monomer-NP interaction on chain conformation.  In theory, the self-consistent polymer reference interaction site model (SC-PRISM) has been employed to calculate the conformation of chains in nanocomposites, revealing that attractive nanoparticles lead to an increase in chain dimension. \cite{Frischknecht2010}  Using molecular dynamics simulations (MD), Starr et al. found that attractive NP fillers caused a 60\% compression in the component of the radius of gyration of chain perpendicular to the NP surface, while the overall radius of gyration increased by 17\%. \cite{Starr2001} At $R_g < R_n$, Smith et al. simulated a mixture of attractive NPs and polymers and found that chain dimension remained almost unchanged. \cite{Smith2002JCP} In contrast, Karatrantos et al. discovered that attractive NPs caused a 20\% expansion in chain dimension, while repulsive NPs had little effect. \cite{Karatrantos2015}  More recently, MD simulations found that NPs with the same diameter as the monomers induced a more compressed chain.  When the ratio of NP diameter to monomer diameter reached 7, the chain expanded by over 7\% at a volume fraction of NPs of 0.4. \cite{Sorichetti2018}

Some Monte Carlo simulations reported the impact of NPs on the chain length of polymers at different size ratios, $R_g/R_n$.  Erguney et al. utilized intramolecular crosslinked potential to prepare collapsed chains, which were regarded as NPs.  Upon the addition of the NPs to macromolecular liquids, they observed cases of chain compression ($R_g>R_n$) and extension ($R_g<R_n$). \cite{Erguney2006,Erguney2008} Some work reported that the conformational distribution of chains remains unchanged regardless of the NP size and the monomer-NP interactions. \cite{Dionne2005} Due to the presence of multiple spatial and temporal scales, as well as the complex effects of exclusion volume and geometric constraints in nanocomposites, there has been inconclusive regarding the variation of polymer dimensions and conformations in extensive simulations.  \cite{Vacatello2001,Vacatello2002,Huang2006,Vogiatzis2011,Ndoro2011}

In macromolecular liquids, the addition of NPs results in the occupation of the space surrounding the tagged chain, which is referred to as the dilution effect. \cite{Li2012,Yamamoto2013}  Recent studies found that in entangled polymer melts, this effect alters the spatial confinement of the chains, leading to a crossover from chain entanglement to NP entanglement. \cite{Li2012,Li2014} The changes in the conformation and motion of tagged chains in nanocomposites are the result of the combined effect of NP confinement and chain confinement.
However, to the best of the authors knowledge, there are no previous reports of the differences in conformations of the tagged chains separatedly influenced by the two different solvent molecules. { In this article, we investigate the conformational properties and the static structures of the tagged chains in a solvent of chain molecules or bead solvents, respectively.  Our main aim is to elucidate the effect of chain connectivity in solvent molecules on the polymer conformations.  }

\section{Model and Simulation Method}\label{sec:2}
\begin{figure}[!t]
\centering
	\includegraphics[width=1\linewidth]{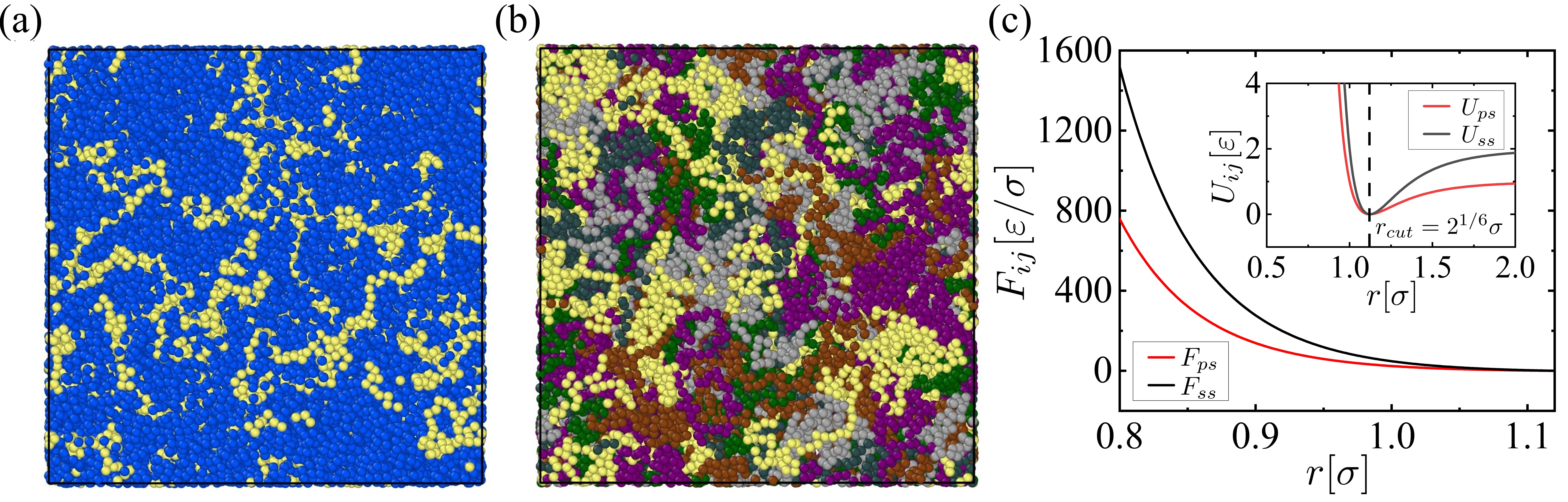}
	\caption{(a) Schematic of the tagged chains immersed in spherical bead solvent molecules. (yellow spheres: the monomers of the tagged chains; blue spheres: beads) (b) Schematic of the tagged chains immersed in linear polymer fluids.  Except for yellow, the other colors represent the chain solvent molecules. (c) The interaction force between tagged monomer and solvent molecule (red line, $\epsilon_{ps} = \epsilon$), and the interaction between solvent monomers (black line, $\epsilon_{ss} = 2\epsilon$). Here, the tagged chain is labelled as $p$, the index $s$ is labelled as $B$ for bead, and is labeled as $p'$ for the solvent chain, respectively. Inset: the solvent-tagged chain interaction potential and solvent-solvent interaction potential. }
\label{fig:1}
\end{figure}
{ We have performed molecular dynamics simulations to study two systems: tagged chains immersed in a solvent of small bead molecules and tagged chains immersed in a solvent of chain molecules as shown in Fig. \ref{fig:1}. In the article, the tagged chain is labelled as $p$, bead molecule is labelled as $B$, and the chain molecule is labelled as $p'$. }

The tagged chains are described as the well-known Kremer-Grest bead-spring model polymers.  \cite{Kremer1990}
The interactions between all particles in our simulations are modeled using the standard Lennard-Jones (LJ) pair potential with a cutoff radius of $r_{cut}=2^{1/6}\sigma$,
{
\begin{eqnarray}\label{1}
U_{ij} (r)=\left\{
\begin{array}{lll}
 4\epsilon\Big[ \Big(\frac{\sigma}{r}\Big)^{12}-\Big(\frac{\sigma}{r}\Big)^6\Big]+\epsilon_{ij},\ r\le r_{cut} \\
 0, \qquad \qquad  \qquad\qquad \ \ \ \ otherwise
\end{array}
\right.
\end{eqnarray}}
where $\epsilon$ and $\sigma$ are the units of energy and length, respectively. {$\epsilon_{ij}$ ensures $U_{ij}(r=r_{cut})=0$}. We set $m$ as the mass of monomer.  The units of temperature, pressure, and time are normalized by $\epsilon/k_B$, $\epsilon/\sigma^3$, and $\sigma(m/\epsilon)^{1/2}$, respectively.

All bonded interactions are described by the finitely extensible nonlinear elastic (FENE) potential
\begin{equation}\label{2}
   U^{B}_{ij} (r)=-\frac{K}{2}R_0^2ln\Big[ 1-\Big(\frac{r}{R_0}\Big)^2\Big]
\end{equation}
where $R_0=1.5\sigma$ and $K=30\epsilon/\sigma^2$.  These parameters precisely determine the location of the minimum of the bond potential at $l_b\approx 0.9606\sigma$.  {The mean bond length in our simulation is $\langle l_b\rangle=0.971$.

For the tagged chains in bead solvent molecules, the interactions between monomers and beads and that between two beads are repulsive and are also represented by the LJ pair potential, where $\epsilon_{pB}=\epsilon$ and $\epsilon_{BB}=2\epsilon$, as shown in Fig. \ref{fig:1}c.  Similarly, for the system of the tagged chains in chain solvent, the interaction between monomers of the tagged chains and monomers of the solvent polymers also follows the LJ pair potential, where $\epsilon_{pp}=\epsilon$, $\epsilon_{pp'}=\epsilon$ and $\epsilon_{p'p'}=2\epsilon$.  These parameters ensure that the repulsive force between solvent molecules is greater than that between the tagged chains and solvent molecules as shown in Fig. \ref{fig:1}c.  Tagged chains are therefore solvophilic.  The diameter of the bead molecule, denoted as $\sigma$, is the same as that of the monomer in the chain molecules. Consequently, the only distinction between the two systems arises from the chain connectivity in the solvent molecules.

In this study, the number of tagged chains is $M_p=40$.  The number of monomers per tagged chain is set as $N_p=100$.  The volume fraction of the tagged chain monomer is defined as $\phi_p=\pi\sigma^3M_pN_p/6V$, which remains constant throughout our simulation, $\phi_p=2.5\times 10^{-2}$.  This is  less than the estimated overlap volume fraction of $\phi_p^*=2.98\times10^{-2}$ based on the radius of gyration at infinite dilution. \cite{Sorichetti2018}  Therefore, the interaction between tagged chains can be neglected.  We define the volume fraction of solvent molecules as $\phi_s=\pi\sigma^3M_sN_s/6V$, where $M_s$ is the number of solvent molecules and $N_s$ is the number of monomers with the same solvent molecules.  In bead solvents, the index $s$ is represented by the symbol $B$. In chain solvents, the index $s$ is represented by the symbol $p'$.  Bead solvent molecules correspond to $N_s=1$.  We investigate the situation with different $N_s$ ranging from 1 to 100. In the article, we primarily report the case with $N_s=100$ as shown in Fig. \ref{fig:1}b. Our simulations focus on the conformations of the tagged chains in the bead and chain solvent molecules at different solvent volume fractions.   }

We carry out all the simulations using LAMMPS software. \cite{Plimpton1995}  Initially, the configurations are prepared by randomly placing the tagged chains and bead molecules in the cubic box.  We use a soft potential to remove the overlaps.  Afterwards, when the solvent volume fraction $\phi_s<0.45$, we perform an equilibration run in NVT ensemble using the Langevin thermostat,
{
\begin{equation}\label{2-2}
  m\frac{d^2\bm{r}}{dt^2}=-\nabla U(\bm{r})-m\Gamma\frac{d\bm{r}}{dt}+\sqrt{2m\Gamma k_BT}\bm{\zeta}(t)
\end{equation}
where $\bm{\zeta}$ represents random force.} The damping coefficient $\Gamma$ is 0.1, temperature is set as $T=1.0$ and integration time step is $\delta t=0.008$.  For the systems of $\phi_s\geq 0.45$, integration time step is $\delta t=0.005$. {The timestep both ensure simulation stability and computational efficiency. } The initial simulation box is much larger than the chain dimension studied. {The 'fix deform' command in LAMMPS is used to shrink the box to a predetermined value, which guarantees the tagged chain volume fraction of $\phi_p=2.5\times 10^{-2}$. Subsequently, we switch to NVT ensemble.}
 During the NVT simulations, the length of  the equilibration run is $2\times 10^8 \delta t$.

\section{Results and Discussion}
\subsection{Radius of gyration}
\begin{figure}[!t]
\centering
	\includegraphics[width=0.8\linewidth]{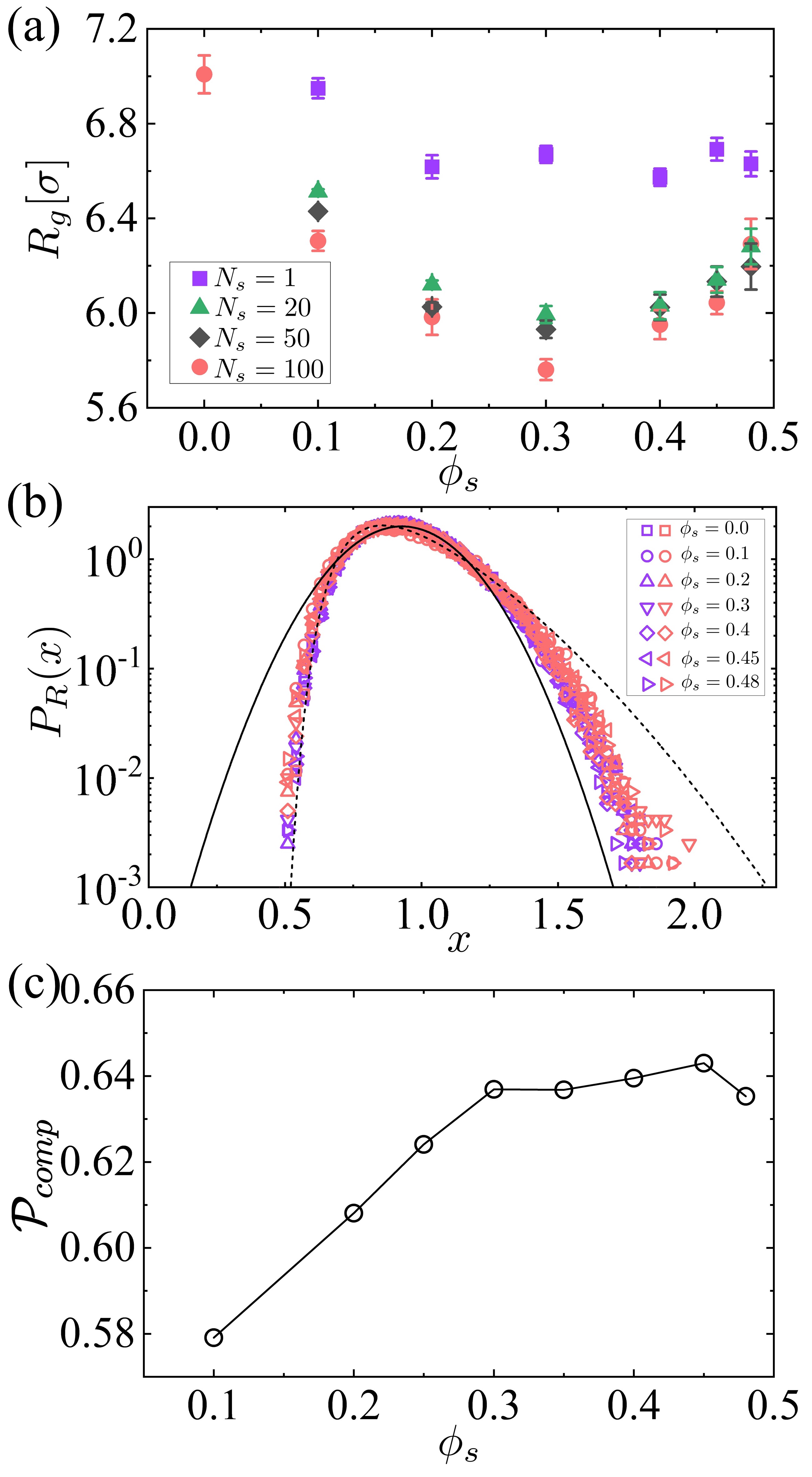}
	\caption{(a) Radii of gyration of the tagged chains for the system of  bead solvent (purple squares) and chain solvent (red circles), respectively. The solvent-free value of $R_g$ is about 7.  (b) Normalized probability distributions of $x=\sqrt{R_g^2/\langle R_g^2\rangle}$ at the several volume fractions.  The solid line is a fit to the Gaussian function.  The dashed line is a fit to Lhuillier's eqn \ref{4} with $b_1=0.2$, $b_2=1.92$, and $C_L=14.31$. (c) The proportion of compressed subchains in a tagged chain as a function of solvent volume fraction.}
\label{fig:2}
\end{figure}
\begin{figure}[!h]
\centering
	\includegraphics[width=0.92\linewidth]{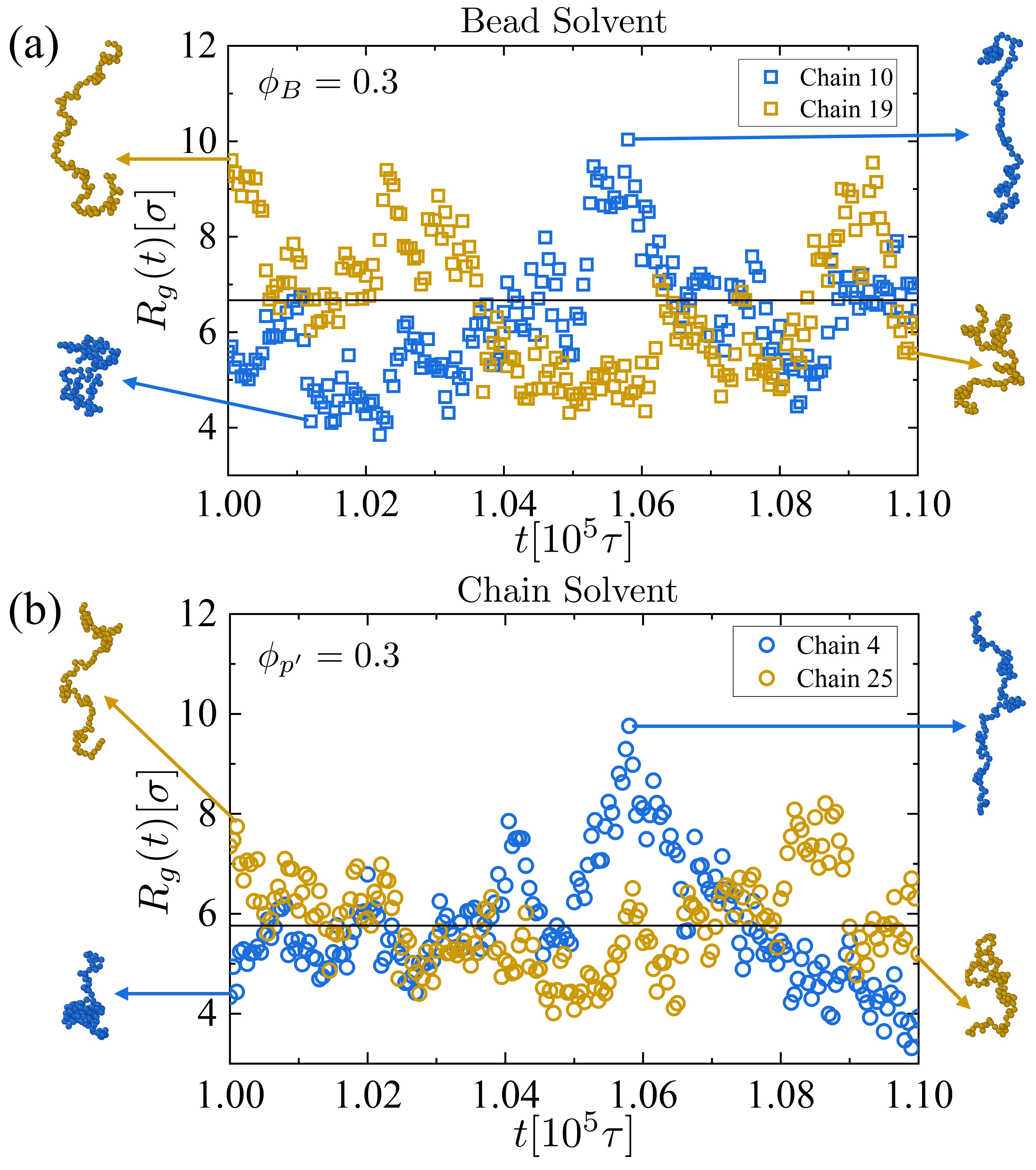}
	\caption{(a) Radius of gyration for two individual tagged chains in bead solvent molecules with $\phi_B=0.3$ as a function of time. (b) Radius of gyration for two individual tagged chains in chain solvent with $\phi_{p'}=0.3$ as a function of time.  Some typical instantaneous configurations are also drawn.}
\label{fig:3}
\end{figure}

We first focus on the tagged chain size in bead solvent and chain solvent by calculating the radius of gyration of a chain, which is given by
\begin{equation}\label{3}
   \langle R_g^2(N_p)\rangle=\langle \sum_{i=1}^{N_p}(\bm{r}_i-\bm{r}_{cm})^2\rangle
\end{equation}
where $\bm{r}_i$ is the position of $i$th monomer belong to the same tagged chains, $\bm{r}_{cm}=\frac{1}{N_p}\sum_{i=1}^{N_p} \bm{r}_i$ is the center of mass of the chain.  $\langle …\rangle$ is calculated by averaging over 1000 independent configurations and 40 tagged chains.

{ Fig. \ref{fig:2}a presents the results for the radii of gyration of the tagged chains in both bead molecules and chain molecules at various solvent volume fractions.  With increasing the concentration of small bead solvent molecules, the tagged chains are compressed and the radius of gyration decreases to about 6.5 at $\phi_B=0.2$ and then fluctuates around the value at higher volume fraction.
}

\begin{figure*}[!t]
\centering
	\includegraphics[width=0.8\linewidth]{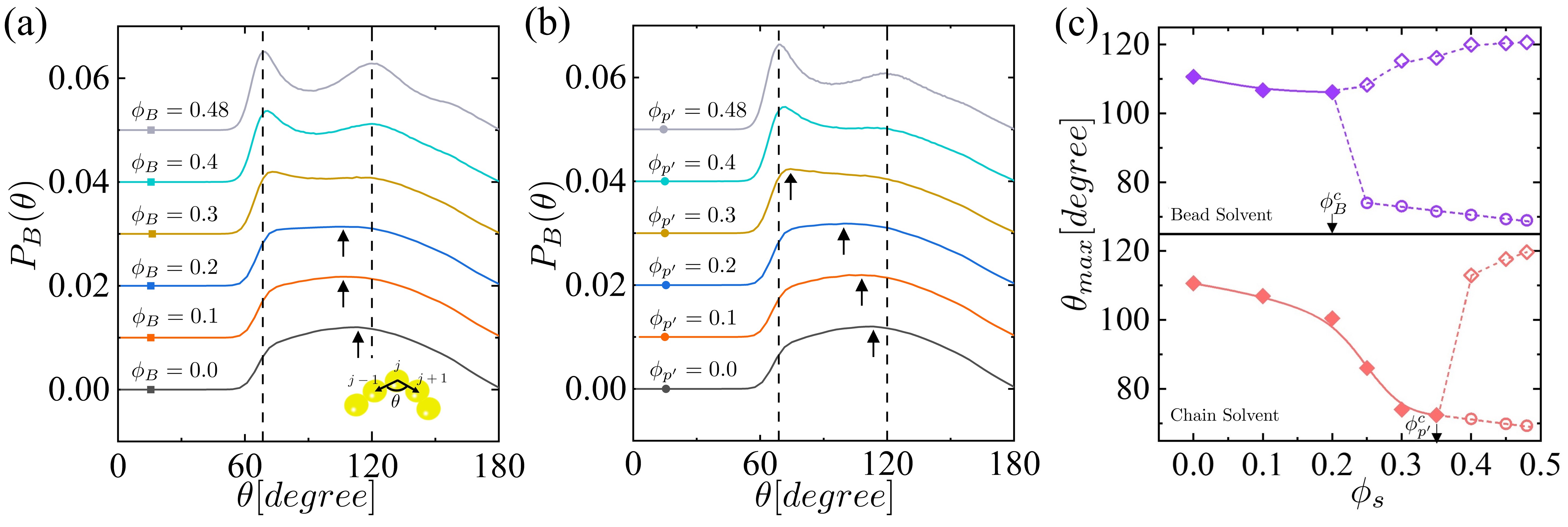}
	\caption{(a) The probability distribution of the bond angles at the several volume fractions for the tagged chains in bead solvent (squares+solid line). (b) The probability distribution of the bond angles at the several volume fractions for the tagged chains in chain solvent (Right: circles+solid line).  (c) The maximum of the bonds of the tagged chains in bead solvent (upper figure) and chain solvent (bottom figure), respectively.  The solid diamonds indicate the unimodal distributions below the crossover volume fraction, $\phi^c$.  The hollow diamonds indicate the peak at obtuse angle.  The hollow circles indicate the peak at acute angle. }
\label{fig:4}
\end{figure*}

On the other hand, in the solvent of chain molecules, the tagged polymer are compressed by a larger extent throughout all volume fractions studied.  {At the polymerization of $N_s=100$, the radius of gyration reaches a minimum at $\phi_{p'}=0.3$, with the tagged chain being compressed by 17.8\% relative to its free chain length.  As $ \phi_{p'}\geq0.3$, the tagged chain gradually expands, reaching a value of $R_g\approx6.3$ at $\phi_{p'}=0.48$.  The qualitative non-monotonic behavior in $R_g$ has also been observed in shorter solvent chains as well.
}

To further investigate the impact of bead and chain molecules on the conformations of tagged chains, the probability distributions of the radius of gyration ($R_g$) as a function of $x=\sqrt{R_g^2/\langle R_g^2\rangle}$ are presented in Fig. \ref{fig:2}b.  In both systems, the normalized probability distributions at different volume fractions collapse onto a master curve, except for high $R_g$ values, suggesting that $P_R(x)$ does not undergo significant changes with the presence of solvent molecules.  Furthermore, $P_R(x)$ shows  marked deviation from a normal Gaussian distribution.   {The reduced chi-square $\chi$ is used to quantify the similarity index.  The Gaussian fit has a reduced chi-square, $\chi>0.0088$ for these systems. }  In this regard, Lhuillier proposed an expression for the probability distribution function of the radius of gyration of an ideal chain in a good solvent as follows: \cite{Lhuillier}
\begin{equation}\label{4}
  P_L(x) = C_Lexp(-b_1x^{-\alpha d}-b_2x^{\delta})
\end{equation}
where $b_1$ and $b_2$ are non-universal constants, $C_L$ is the normalized factor.  The scaling exponents $\alpha = (\nu d-1)^{-1}$ and $\delta =(1-\nu)^{-1}$, where $d$ is space dimension and $\nu=1/2$ for ideal chain.

{  In the bead solvent, $\chi=0.0036$, whereas for chain solvent molecules, it stands at $\chi=0.0033$. As shown in Fig. \ref{fig:2}, the fitting $P_R(x)$ demonstrates excellent agreement with Lhuillier's equation, particularly near or for small $x$ values ($x\leq 1.3$), suggesting that the conformation of the tagged chain remains expanded, resembling that seen in a good solvent solution, irrespective of whether it is in a small bead molecule solvent or a chain solvent.

Furthermore, we have introduced a quantity to measure the ratio of compressed subchains within a tagged polymer chain. The tagged chain, consisting of 100 monomers, is divided into 10 subchains, each comprising 10 monomers. We calculate the $R_g$ of these subchains in the solvent molecules of the chain. If the $R_g$ value is smaller than the corresponding value in the solvent of small bead molecules, we classify the subchain as compressed.  Through this method, we can evaluate the proportion of compressed subchains within a tagged chain, $\mathcal{P}_{comp}$. As shown in Fig. \ref{fig:2}c , the proportion increases linearly at $\phi_s\leq0.3$.  At higher solvent density, the saturation of the curve implies that the proportion of compressed subchains in a tagged chain remains constant, consistent with the physical picture mentioned above.
}

We also plot the radius of gyration of some specific tagged chains against time in Fig. \ref{fig:3}.   Two chains were selected at $t=10^5\tau$, with one exhibiting a higher value and the other a lower value compared to the average chain size.   It is important to note that the chain sizes change dynamically over time, and the specific chain undergoes alternating cycles of expansion and compression.  For example, chain 10 initially shows a compressed conformation. However, about $4\times 10^3\tau$ later, the radius of gyration of chain 10 surpasses the average value, indicating expansion of the chain.

\subsection{Bond angles}
{In general, due to internal torsion energy or anisotropy of chain stretching under shear, chains with the same $R_g$ may quit different bond angle distributions.\cite{Kroger2022,Kroger1997}In the linear KG polymer model, The conformation and size of chains are believed to be closely connected to the average value and distribution of bond angles. \cite{Kob2020}  }

Fig. \ref{fig:4}b exhibits the bond angle distribution functions of the tagged chains at various volume fractions in bead and chain solvent, respectively.  At low volume fractions, the bond angle distribution displays a single peak around $\theta\approx 110^\circ$, indicating a expanded conformation for the tagged chains.  As the volume fraction of the solvent increases to a crossover value ($\phi^c_s$), the distributions transition into bimodal distributions with narrower peaks.  The position of the peak at low volume fractions shifts towards smaller acute angles and larger obtuse angles above the crossover volume fraction of the solvent. Fig. \ref{fig:4}c illustrates the bifurcation of peak values in the distribution function, $P_B(\theta)$, as a function of volume fraction. The crossover volume fractions are $\phi^c_B=0.2$ for bead solvent molecules and $\phi^c_{p’}=0.35$ for chain solvent molecules, respectively. The position of this peak ranges from 110 degrees at $\phi_s=0$ to approximately 70 degrees and 120 degrees at $\phi_s=0.48$. { These two saturated bond angles closely resemble the bond angles found in an FCC lattice, with minor discrepancies attributed to the coexistence of two competing intrinsic lengths in the model polymer, bond length and monomer diameter. A decrease in the mean bond angle is correlated with a corresponding decrease in the mean bond length.

It is worth noting that in solvents consisting of small beads, below the crossover volume fraction, the angular distribution exhibits a prominent obtuse peak, while above this threshold, additional branches with acute angles become apparent. Conversely, in solvents composed of chain molecules, below the crossover volume fraction, the angular distribution already displays compressed peaks with acute angles. As the density of solvent molecules increases further, an obtuse angle branch emerges. These findings indicate a physical phenomenon where the chain connectivity in solvent molecules leads to compression of the tagged chains. When surpassing the crossover density, the presence of obtuse angle branches signifies stretching of certain chains and consequently results in an increase in $R_g$. This bifurcation suggests that as the solvent volume fraction increases, a portion of a tagged chains undergoes compression while another portion experiences stretching.}

\begin{figure}[!t]
\centering
	\includegraphics[width=0.8\linewidth]{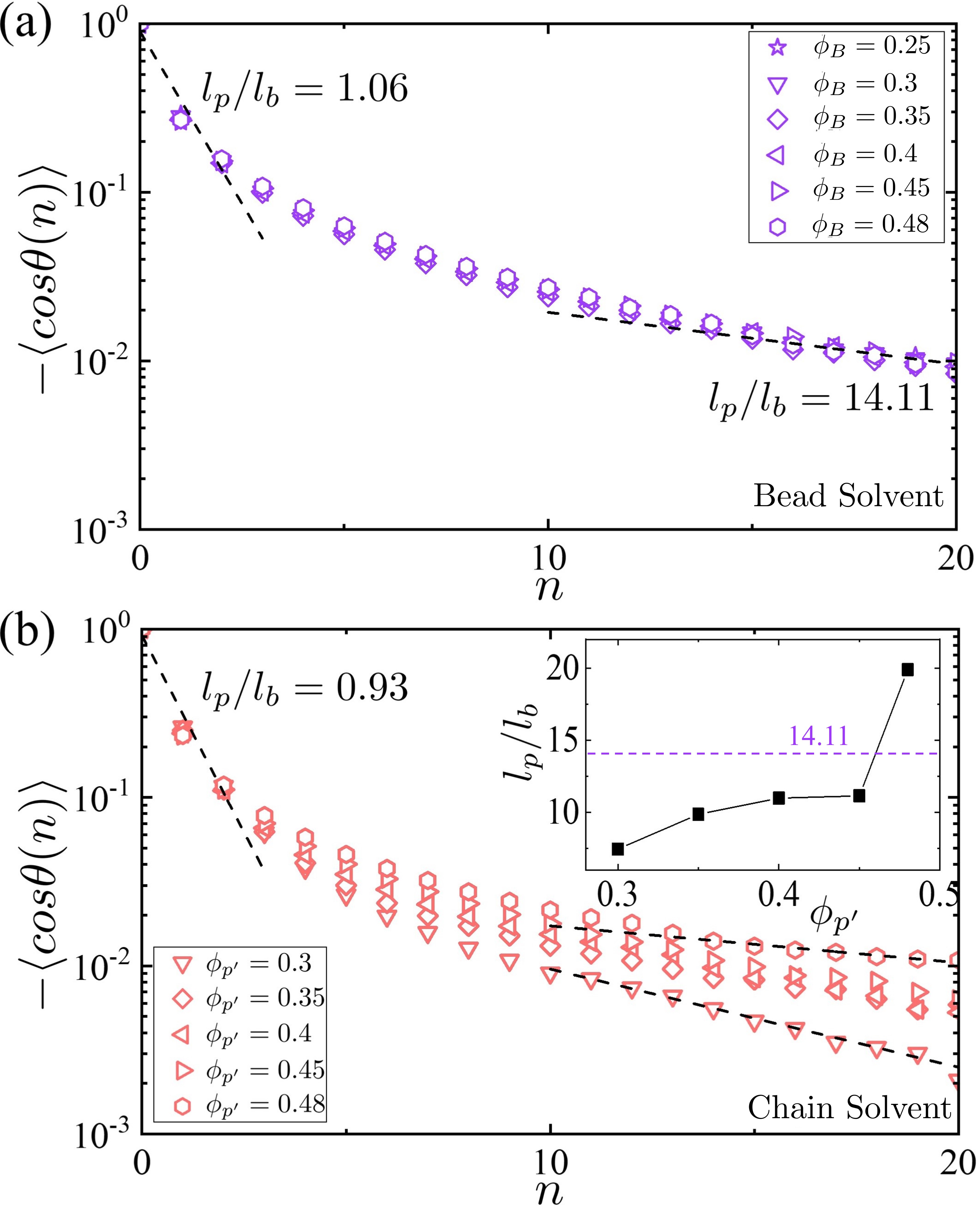}
	\caption{(a) The correlation function between two bonds along a identical tagged chain in bead solvent as a function of contour distance.  (b) The correlation function between two bonds along a identical tagged chain in chain solvent as a function of contour distance.  The persistent length can be extracted from the exponential decay function, $exp(-nl_b/l_p)$ at short distance and long distance, respectively.  Inset: the persistence length as a function of volume fraction of the solvent polymer, $\phi_{p'}$.
}
\label{fig:5}
\end{figure}

To characterize the impact of solvent addition on the persistent length of the tagged chains, we analyze the correlation function $-\langle cos \theta(n)\rangle$ between two bonds along the same chain, where $n$ represents the contour distance. In the case of a freely rotating chain (FRC) model with a bond length $l_b$, the correlation function exhibits exponential decay with increasing distance $n$, as depicted by $-\langle cos \theta(n)\rangle=exp(-nl_b/l_p)$. Interestingly, when the volume fraction of beads exceeds the crossover density, $\phi_B^c=0.25$, the correlation function remains mostly unchanged, as illustrated in Fig. \ref{fig:5}a. The linear-log plot clearly illustrates the exponential decay of $-\langle cos \theta(n)\rangle$ at both small and large contour distances. However, at intermediate contour distances, the bond-bond correlation deviates from the exponential decay predicted by the FRC model. The persistence lengths extracted from the initial and final decays of the function are $l_p/l_b=1.06$ and $l_p/l_b=14.11$, respectively.

In chain solvent with varying volume fractions, the bond-bond correlation functions of the tagged chains remain unchanged at small contour distances.  As the contour distances increase, the decay of $-\langle cos \theta(n)\rangle$ becomes slower at higher volume fractions as shown in the inset of Fig. \ref{fig:5}b.   { We also calculate the bond correlation function of the tagged chain in a solvent consisting of short chains, where the polymerization degree $N_s=20$ is smaller than the typical entanglement length of the system, $N_e\approx25$, as estimated by primitive path analysis at the highest volume fraction we studied.  Despite this, we still observe a slowdown in the decay of $-\langle cos \theta(n)\rangle$ at higher volume fractions.  We  suggest that this phenomenon is caused by the confinement imposed by the chain solvent.}

Moreover, in the volume fraction range of 0.3 to 0.45, the persistence lengths extracted from the correlation functions in polymer fluids are found to be smaller than that in bead solvent.  However, at an even higher volume fraction of $\phi_s=0.48$, the persistence length in chain solvent exceeds that in bead solvent. This implies that at extremely high densities, polymer chains in the fluid exhibit even greater rigidity and resistance to conformational changes compared to bead fluids.

\subsection{Tagged chain structure and effective chain-solvent interactions}
\begin{figure}[!b]
\centering
	\includegraphics[width=0.8\linewidth]{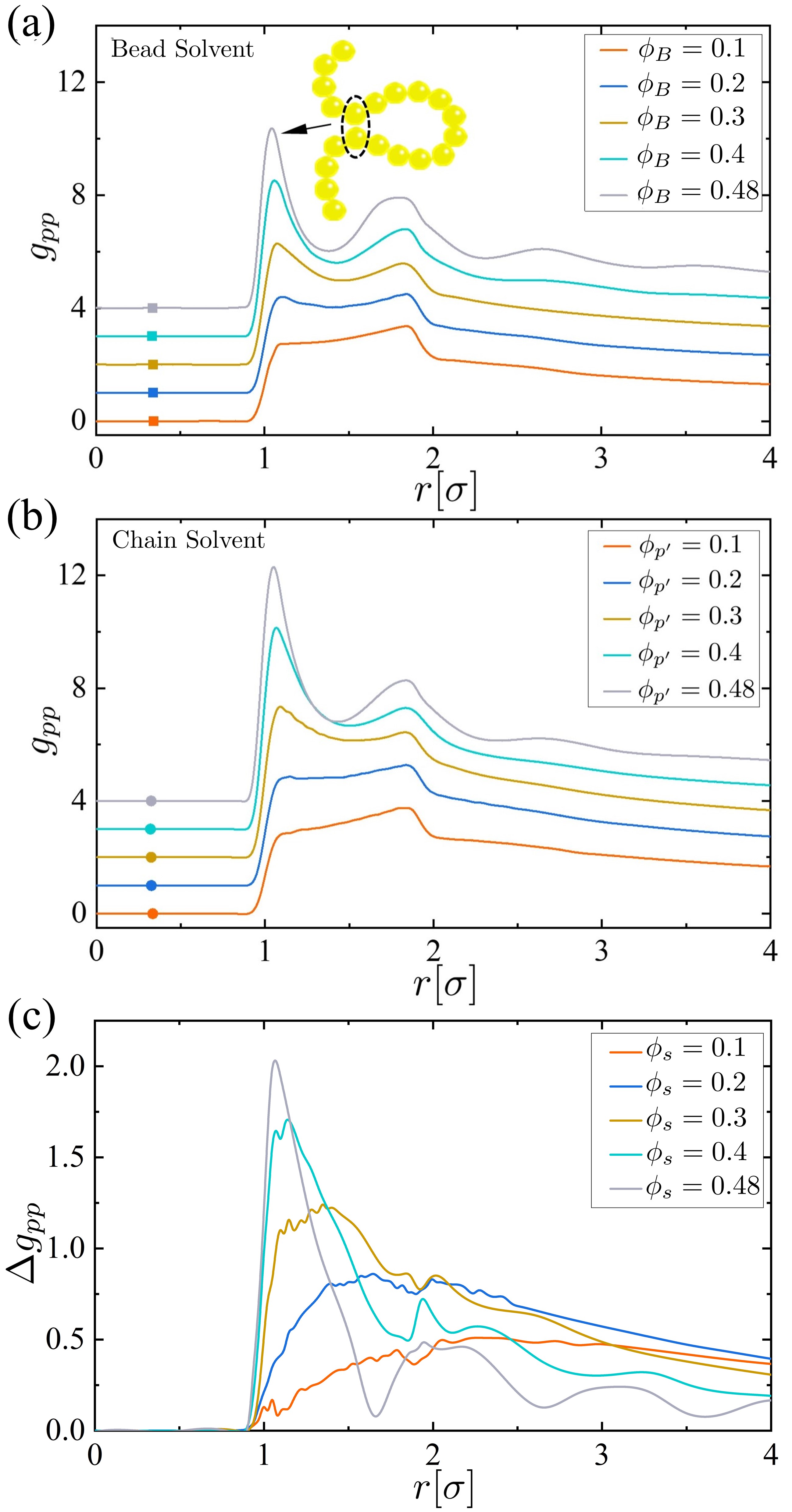}
	\caption{(a) The radial distribution function of the tagged chains in the bead solvent at the several volume fractions.  The functions are shifted upward for clarity.  The main peak indicates the nonbonded contact in the identical chain as shown in the cartoon.  (b) The modified radial distribution function of the tagged chains in the chain solvent at the several volume fractions. (c) The RDF difference function $\Delta g_{pp}=g_{pp}^{p'}(r)-g_{pp}^B(r)$.}
\label{fig:6}
\end{figure}
\begin{figure}[!t]
\centering
	\includegraphics[width=0.8\linewidth]{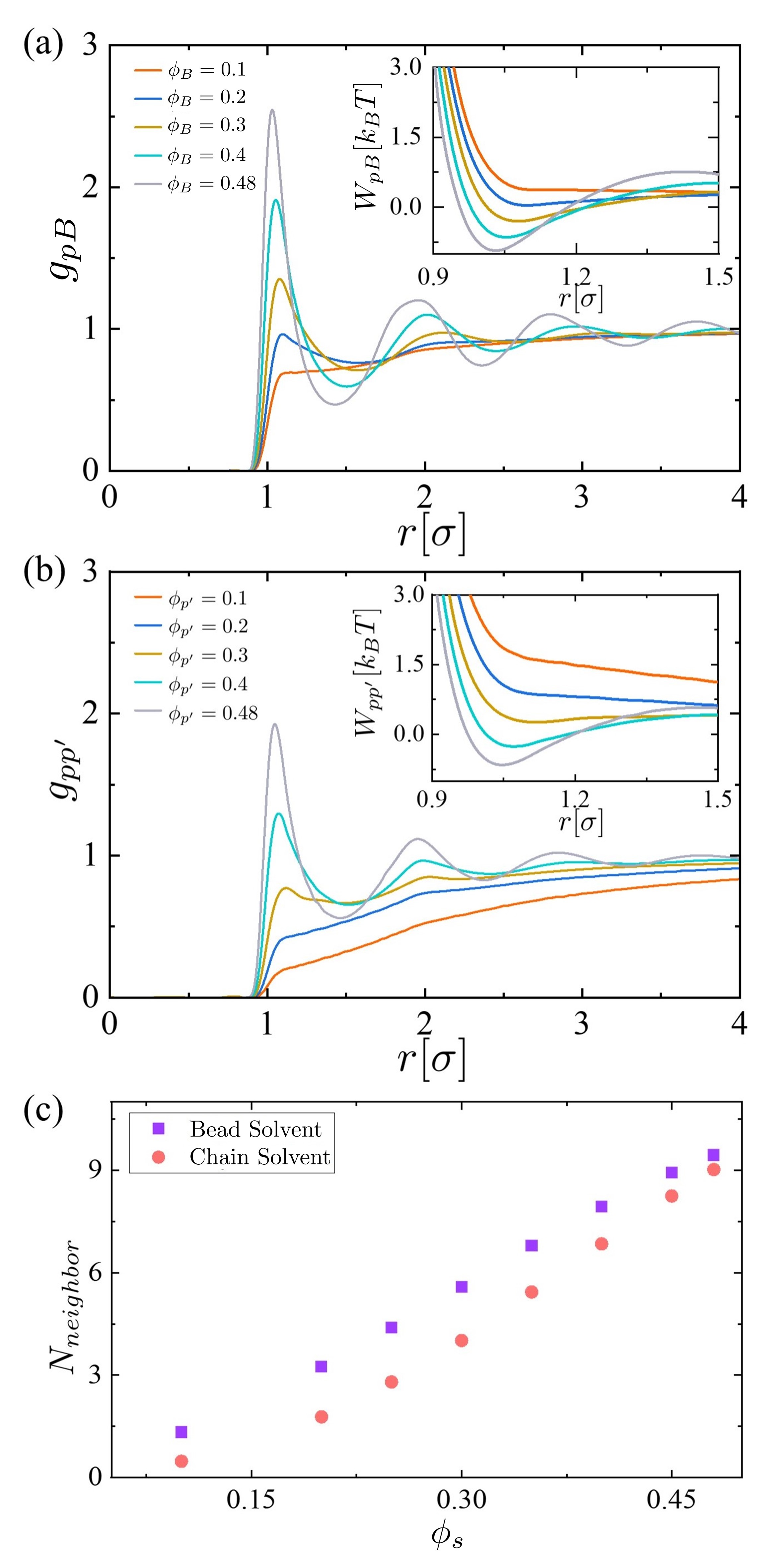}
	\caption{(a) The radial distribution functions of the bead solvent and the monomer of the tagged chains.  (b) The radial distribution functions of the solvent monomer and the tagged monomer.  (c) The number of the solvent molecules occuping the first shell as a function of volume fractions. }
\label{fig:7}
\end{figure}
To further quantify the conformational behaviors of the tagged chains in the two systems, we consider the radial distribution function (RDF) of the tagged chain which is defined as
\begin{equation}\label{5}
   g_{pp}(r)=\frac{1}{\rho_p N_p}\Big\langle \sideset{}{'}\sum_{\substack{i,j=1\\i\neq j}}^{N_p}\delta(r-|\bm{r}_i^p-\bm{r}_j^p|)\Big\rangle
\end{equation}
where $\rho_p= N_pM_p/V$ is the total number density of monomers of the tagged chains.  This $\sum’$ represents the summation over non-bonded monomers $i$ and $j$ in the identical tagged chains.  Neighboring interactions resulting from bonding, at $r=2^{1/6}\sigma$, are excluded from the RDF calculations. In the regime of dilute tagged chains, the RDF primarily reflects the intrachain correlations as the main contribution.

As shown in Fig. \ref{fig:6}, a small addition of bead molecules or chains results in less pronounced first peaks in the RDF, indicating minimal contact interaction between the monomers within the same tagged chain.  This suggests that the conformation of the chain is not significantly compressed.  With an increase in solvent density, the first peak in the RDF starts to develop.  Upon reaching a volume fraction exceeding 0.25, the first peak surpasses the second peak.  This suggests that intrachain contact interactions become more important, leading to a portion of the tagged chain being compressed by the surrounding solvents, as illustrated in the cartoon of Fig. \ref{fig:6}a.  Comparing chain solvent to bead solvent, the intrachain correlation of the tagged chain is larger in chain solvent.  It is characterized by a higher first peak value, implying that chain confinement induces greater compression.  {The difference between the RDFs can be quantified by $\Delta g_{pp}=g_{pp}^{p'}(r)-g_{pp}^B(r)$.  The magnitude of this difference function tends to be larger at higher solvent densities, particularly around $r=1$.}

In order to quantify the effective chain-solvent interaction, we measure the radial distribution function for monomer-solvent pair defined as,
\begin{equation}\label{6}
   g_{ps}(r)=\frac{V}{N_pN_s}\Big\langle \sum_{i=1}^{N_p}\sum_{j=1}^{N_s} \delta(r-|\bm{r}_i^p-\bm{r}_j^s|)\Big\rangle
\end{equation}
where the index $s$ is represented as $B$ for bead fluids, and as $p’$ for polymer fluids.

As shown in Fig. \ref{fig:7}, the strength of monomer-solvent correlation at contact, $r=\sigma$, notably increases with higher solvent loading.  In bead fluids, the main peak at contact becomes evident at $\phi_B=0.2$.  However, in polymer fluids, the main peak remains suppressed until the monomer volume fraction exceeds 0.25.  Physically , this suppression is commonly referred to as the correlation hole effect, which can be attributed to the self-screening of the tagged chains. \cite{Schweizer1990} Within the radius of gyration, a polymer coil inhibits the approach of monomers from different chains, resulting in an effective repulsive interaction. The average number of nearest neighbor particles around the tagged monomers can be estimated by performing an integral $N_{neighbor}=\int_0^{1.5\sigma}\rho_s g_{ps}(r)4\pi r^2 dr$.  As the volume fraction increases from 0.1 to 0.48, the average number of neighboring particles increases from below 2 to greater than 9, as shown in Fig. \ref{fig:7}c.  Due to the correlation hole effect, polymer fluids exhibit a smaller $N_{neighbor}$ than bead fluids at the same volume fraction.

The potential of the mean force (PMF) can be utilized to describe the effective interactions between monomers and solvents, taking into account all enthalpic and entropic effects in the solvent medium. \cite{Lisa2008} The PMF is calculated by,
\begin{equation}\label{7}
W_{ps}(r) = -k_BTln[g_{ps}(r)]
\end{equation}
where $k_B$ is Boltzmann constant.

As shown in the inset of Fig. \ref{fig:7}a and \ref{fig:7}b, at low solvent density, the PMFs exhibit a monotonic decrease with increasing interparticle distance and lack any depletion-like attraction.  This is due to the absence of solvent particles being enriched around the tagged chain, as illustrated in the monomer-solvent RDF.  Nevertheless, as the solvent particle loading increases, an attractive well forms at short distances, resulting in the solvent effectively pulling the particles together.  The depth of this attractive well serves as a quantification of the level of solvent-monomer mixing.  In the case of polymer fluids, the correlation hole effect induces a stronger repulsion between solvent molecules and the tagged monomers.  Consequently, this attractive well appears at higher volume fractions, $\phi_{p’}\geq 0.25$.

\subsection{Pore size distribution}
\begin{figure}[!t]
\centering
	\includegraphics[width=0.8\linewidth]{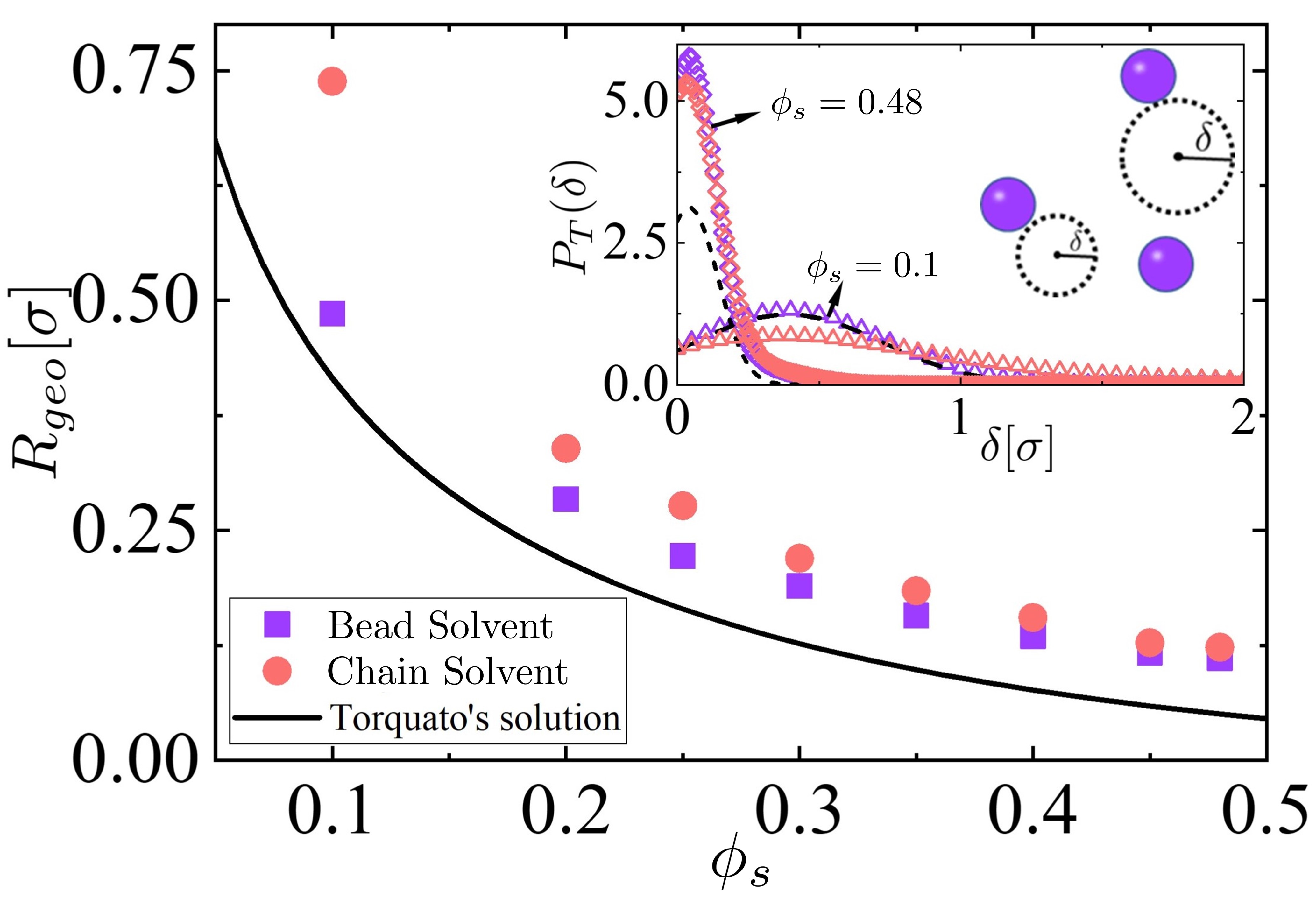}
	\caption{The pore size as a function of volume fractions for the systems of bead fluids (purple squares) and chain fluids (red circles), respectively.  The solid line represents the Torquato's solution. Inset: the pore size distributions are shown at $\phi_s=0.1$ (trangles) and $\phi_s=0.48$(diamonds), respectively.  The short-dashed and long-dashed lines indicate the calculated PSD using eqn \ref{9} at $\phi_s=0.1$ and $\phi_s=0.48$, respectively. The cartoon shows the algorithm of Torquato’s PSD. }
\label{fig:8}
\end{figure}
In our study, we explore the influence of beads and polymer chains on the geometrical confinement, which can be characterized by the pore size.\cite{Li2012, Sorichetti2018}  In a solution, there are two distinct regions: one occupied by the solvent and the other being a void region. The size of the void region quanfities the geometrical confinement influenced by the presence of solvent particles.  To measure this, we calculate the pore size distribution (PSD) function, originally developed for solid porous media, which has been recently generalized and successfully applied to estimate the geometrical mesh size in polymer melts. \cite{Kob2020}

Following the algorithm of the PSD proposed by Torquato, \cite{Torquato2002} we begin by randomly placing points in the void region. Subsequently, we determine the interface distances, $\delta$, between these points and their nearest monomers. Through repeated measurements of a large number of the distances, we obtain the PSDs.  The mean pore radius can then be obtained by evaluating the following integral.

\begin{equation}\label{8}
R_{geo} = \int_0^{\infty} \delta P_T(\delta)d\delta
\end{equation}

In addition, an analytical expression for the PSD $P_T(r)$ of hard spheres has been also obtained by Torquato et al.,
\begin{eqnarray}\label{9}\nonumber
P_T^a(r, \phi_s)&=&24\phi_s(1-\phi_s)(er^2+fr+g)\times \\
&&exp(-\phi_s(8er^3+12fr^2+24gr+h))
\end{eqnarray}
where $r$ is the The distance between particle centers.  $e(\phi_s)$, $f(\phi_s)$,$g(\phi_s)$,and $h(\phi_s)$ are the function of volume fraction can be found in ref. \citenum{Torquato1990} .

The inset of Fig. \ref{fig:8} shows the pore size distribution (PSD) for bead fluids and chain fluids at different solvent densities.  It is observed that the peak intensity of the PSD increases, while the peak position decreases, with increasing solvent loading. This can be attributed to the stronger geometric constraints induced by higher solvent loading, which leads to the occupation of larger portions of the void region.  In polymer fluids, the chain connectivity plays a crucial role in the monomer dispersion, consequently influencing the distribution of the void region.  As shown in the inset of Fig. \ref{fig:8}, the fluctuation in pore size becomes more pronounced, resulting in a broader width of the PSD function.  Moreover, the calculated PSD of bead fluids at low density shows good agreement with the analytical PSD, $P_T^a(r)$.  However, at high density, it is observed that the analytical expression, eqn \ref{9}, underestimates the PSD compared to the simulation results. {While this analytical expression cannot perfectly capture the magnitude of the simulation PSD at high densities, it successfully predicts the peak location, which directly determines the pore size.  Therefore, the geometrical constraint $R_{geo}$ calculated by the Torquato's equation are no significant deviation from the simulation results at high density as shown in Fig. \ref{fig:8}.
}

With an increase in solvent loading, the pore size experiences a substantial reduction. At $\phi=0.48$, the pore size $R_{geo}$  decreases to approximately $ 0.12\sigma$. Comparatively, chain solvents exhibit larger pore sizes than bead solvent across the range of volume fractions studied. However, this difference becomes smaller as the density increases.  At higher densities, both systems demonstrate similar particle dispersion, leading to a similar pore size as shown in the Fig. \ref{fig:8}.

\section{Conclusion}
Using molecular dynamics (MD) simulations, we compare the effects of bead molecule confinement and chain confinement on the conformational behaviors and static properties of the tagged chains for the first time.  Firstly, we show that chain confinement resulted in a greater degree of compression compared to bead molecule confinement. Additionally, in polymer fluids, we identify a minimum radius of gyration at a monomer volume fraction of $\phi_{p'}=0.30$.  It is worth noting that the probability distributions of chain size remain consistent across different solvent densities, regardless of whether the solvent is composed of bead molecules or chain molecules.  Furthermore, we found that the chain compression is accompanied by a decrease in the average bond angle.  As the solvent density increases, the distribution of bond angles displayed a crossover from a unimodal to a bimodal function. This indicates the presence of both compressed and expanded regions within the chain.  Notably, the crossover volume fraction for polymer fluids is larger than that for bead fluids.  At high solvent density, the bond orientation correlation functions along an identical chain in bead solvent remain unchanged across different $\phi_B$.  On the other hand, in polymer fluids, the functions exhibit notable changes with increasing chain molecule loading.  In particular, a longer persistent length of the tagged chain is obtained, which can be extracted from the bond orientation correlation function at long contour distance.

Afterwards, we study the radial distribution function (RDF) for the tagged chains in both bead solvent and chain solvent.  Our findings reveal that the main peak of the RDF increased as the solvent density increases.  In particular, the larger value of the main peak in polymer fluids indicate that chain confinement induces a greater extent of compression for the tagged chain compared to bead confinement.  By calculating the partial RDF, $g_{ps}$, we are able to extract the solvent-chain effective interaction and the number of nearest neighbors.  Our results show that as the solvent density increases, there is an observable effective attraction between the solvent and the chain.  In polymer fluids, the correlation hole effect lead to a weakening of this attractive interaction and a decrease in the number of nearest neighbors around the tagged monomers.  Finally, our calculations suggest that chain connectivity induces solvent nonhomogeneity, resulting in a broader pore size distribution and larger pore sizes in the system.

This work provides valuable guidelines for chain conformations and structures by comparing the effects of different solvent molecules. It serves as a foundation for a variety of open problems that warrant future research, such as (i) the scaling physics of a tagged polymer in different solvent molecules.  (ii) the role of the ratio of bead size to monomer size and the presence of attractive beads on chain conformations. (iii) the influence of chain confinement and bead confinement on dynamic quantities, including entangled dynamics at large density and glassy behaviors at low temperature.

\begin{acknowledgments}
This work is supported by National Natural Science Foundation of China (No. 12374218,11904320,11774041,and 12204339), the Fundamental Research Funds for the Central Universities, China under Grant No. SWU-KQ22033 and the Natural Science Foundation of Chongqing, China under Grant No. CSTB2022NSCQ-MSX0512, and the Science and Technology Research Program of Chongqing Municipal Education Commission under Grant No. KJQN202200210. The authors are grateful to the High Performance Computing Center of Southwest University for carrying out the partial numerical calculations in this work on its blade cluster system. 
\end{acknowledgments}

\bibliography{main} 

\end{document}